\documentstyle[aps,epsf]{revtex}
\newcommand{\tr}{\mbox{Tr} }
\newcommand{\ket}[1]{\left | #1 \right \rangle}
\newcommand{\bra}[1]{\left \langle #1 \right |}
\newcommand{\proj}[1]{\ket{#1} \! \bra{#1}}

\begin{document}
\title{Noise effects for the Depolarizing Channel}
\author{
Julian Juhi-Lian Ting
}
\address{jlting@multimania.com}
\date{\today}
\maketitle
\begin{abstract}
The possibility of stochastic resonance of a quantum channel and
hence the noise enhanced capacity of the channel is
explored by considering the depolarizing channel.
The fidelity of the channel is also considered.
Although there is no clear evidence for noise enhanced capacity found,
there is enhancement for the fidelity for the depolarizing channel.

\end{abstract}

\section{Introduction}

Recently, because of the development of quantum computers\cite{AS}
people have become interested
in information transmission thorough quantum channels\cite{S}. 
Quantum information theories can be used to describe processes 
such as data storage, quantum
cryptography\cite{DEJ}, and quantum teleportation\cite{BBP}.
However, after an initial burst of papers following Shor's discovery of
quantum factoring algorithm\cite{SH}, almost every work is aiming to solve
the decoherence problem.
There are people using NMR techniques, which provide longer
decoherence time than previous techniques,
claiming they can built a quantum
computer with a cup of coffee\cite{GC}.
There are also people trying to use various software methods,
in particular, quantum error correcting codes, to correct decoherence
induced errors.
Apparently, decoherence is a hurdle need to be surmounted 
before quantum computers can be 
materialized.
However, is decoherence, the counterpart of classical noise, really nuisance?
For people who know stochastic resonance, the answer is perhaps 'no'.

Stochastic resonances in nonlinear dynamics is about
noise enhancement of some useful properties of a system.
It have been considered for
both quantum and classical periodical signals\cite{BG}. The consideration
for aperiodical signal cases is rather recent\cite{BZ,FCB}.
However, most works
are concerned about classical channels. Therefore, the consideration
for quantum channels is of interest in itself whether
there is noise enhancement or not, because it is a missing piece in the 
theories of stochastic resonance.

In the periodical cases spectral properties of a system are generally used
to characterize their performance.
However, it has been recognized recently that spectral properties of a system
are adequate for describing 
the system only when the system is linear\cite{RABI}.
Mutual information, a kind of entropy change in statistical mechanics, seems
to be an even better parameter to describe the resonance.
In the works for classical aperiodical stochastic resonances the mutual
information between the input and the output of the channel has been used
as a measure of channel performance to be enhanced by the noise.
A peak in the mutual information versus noise curve indicate resonance.
To study the noise enhanced channel capacities of quantum channels one
need a measure for noise and a measure for the channel capacities, or
a measure for any other property interested.
The problem is: which quantity can be
used as the {\em correct} measure ?

In what follows the capacity and fidelity versus noise relation of the
depolarizing   channel \cite{SS} is
considered.
The capacity versus flipping rate relation
of the depolarizing channel has been previously considered by
Lloyd\cite{LL}.
However, it is the capacity versus noise relation, which
has its interpretation in stochastic resonance, is of more importance
in physics. 
A similar consideration has been presented for the
two-Pauli channels by Ting\cite{T}.
This paper therefore has three folds of purposes:
Firstly, a more through explanation of the problem considered than
the previous paper is taken. Secondly, we tried to establish links
with previous works in quantum computing, 
in particular those related to Lloyd's, and in other fields of physics\cite{JP}.
Thirdly, we tried to explore one more channel.

\section{the Noisy Channel Model}
Schumacher {\it et al.}\cite{S96b} have developed a quantum information
theory. In their formulation
a quantum channel  can be considered as
a process defined by an input density matrix 
$\rho_x$, and an output density matrix $\rho_y$, with the process
described by a quantum operation, ${\cal N}$,
\begin{eqnarray} 
\rho_x \stackrel{{\cal N}}{\rightarrow} \rho_y.
\end{eqnarray}
Because of decoherence, the super-operator $\cal N$
is not unitary.
However,
on a larger quantum system that includes
the environment $E$ of the system, the total evolution
operator $U_{x E}$ become unitary.  This environment may be considered
to be initially in a pure state $\ket{0_{E}}$ without
loss of generality.  In this case, the
super-operator can be written as
\begin{equation}
        {\cal N} (\rho_{x}) = \tr_{E} U_{xE} \left (
                                  \rho_{x} \otimes \proj{0_{E}} \right )
                                  {U_{xE}}^{\dagger} .
\label{channel}
\end{equation}
The partial trace, $\tr_E$, is taken over environmental degree of freedom.
Eq.~(\ref{channel}) can be rewritten as a completely positive linear
transformation acting on the density matrix:
\begin{equation}
{\cal N} (\rho_x)=\sum_i A_i\rho_x A_i^\dagger\;,
\label{AnklesOfHair}
\end{equation}
in which the $A_i$ satisfy the completeness relation
\begin{equation}
\sum_i A_i^\dagger A_i = I\;,
\label{HankThoreau}
\end{equation}
which is equivalent to requiring $\tr [{\cal N} (\rho_x )]=1$.
Conversely, any set of operators $A_i$ satisfying
Eq.~(\ref{HankThoreau}) can be used in Eq.~(\ref{AnklesOfHair})
to give rise to a valid noisy channel in the sense of
Eq.~(\ref{channel}).
Eq.~(\ref{AnklesOfHair}) is in fact a Schr\"odinger evolution of the 
density matrix.

In order to study the noise effects, we certainly need a definition of noise.
Schumacher has also gave a definition of noise, i.e.
\begin{eqnarray}
N (\rho_x,{\cal N}) \equiv - \tr (W \log_2 W), \end{eqnarray}
with
\begin{eqnarray}
W_{ij} \equiv \mbox{Tr}(A_i \rho_x A_j^{\dagger}).
\end{eqnarray}
It measures the amount of information
exchanged between the system $x$ and the environment $E$ during
their interaction\cite{S}.
If the environment is initially in a pure state,
the entropy exchange is just the environment's entropy after the
interaction.

The next quantity we need is a measure of any quantity we are interested in.
For a communication channel the most important quantities are its
capacity and its fidelity.        
Although the coherence information is generally believed to
represent only a lower bound on the channel
capacity in the Shannon's definition, it can be used to represent the channel 
capacity without talking about encoding\cite{BST,BKN}.
It is defined as
\begin{eqnarray}
C (\rho_x,{\cal N}) \equiv H \left(
        {\cal N}(\rho_x) \right) -
        N (\rho_x,{\cal N}),
\end{eqnarray}
in which
$H ( \bullet ) = - \tr \left[ \bullet \log_2 \bullet \right]$ 
is the von Neumann entropy\cite{N}. 
Coherent information plays a role in quantum information theory 
analogous to that played
by the mutual information in classical information theory.
It can be positive, negative, or zero.

\section{The Depolarizing Channel}

A depolarizing channel
can be written in terms of $A_i$'s in Eq.~(\ref{AnklesOfHair}) as

\begin{equation}
A_1=\sqrt{x}\,I\;,\;\;\;\;
A_2=\sqrt{ {\frac{(1-x)}{3} }}\,\sigma_1\;,\;\;\;\;
A_3=\sqrt{ {\frac{(1-x)}{3} }}\,\sigma_3\;,\;\;\;\;
A_4=-i\sqrt{ {\frac{(1-x)}{3} }}\,\sigma_2\;,
\end{equation}
where $I$ is the identity matrix and $\sigma_1$, $\sigma_2$, and
$\sigma_3$ are the Pauli matrices, i.e.,
\begin{equation}
\sigma_1=\left(\begin{array}{cr}
0 & 1\\
1 & 0\end{array}\right),\;\;\;\;
\sigma_2=\left(\begin{array}{cr}
0 & -i\\
i & 0\end{array}\right),\;\;\;\;
\sigma_3=\left(\begin{array}{cr}
1 & 0\\
0 & -1\end{array}\right).
\end{equation}
This channel can be interpreted as:  with probability
$x$, it leaves the qubit alone; with probability $(1-x)/3$ it either
flip the qubit amplitude ($A_2$), or rotate the qubit ($A_3$), 
or do both flip and rotation to the qubit($A_4$). 
In other words, if one has a 
quantum state
\begin{equation}
|\Psi\rangle =
\alpha\,|0\rangle + \beta\,|1\rangle\;,
\end{equation}
goes through such a channel, the resulting change will be
\begin{eqnarray}
\ket{\Psi} &\stackrel{x}\longrightarrow & \ket{\Psi}\;,\nonumber\\
\ket{\Psi} &\stackrel{(1-x)/3}\longrightarrow & \sigma_1\ket{\Psi}=
\alpha\,\ket{1}+\beta\,\ket{0}\;, \nonumber\\
\ket{\Psi} &\stackrel{(1-x)/3}\longrightarrow & \sigma_3\ket{\Psi}=
-\alpha\,\ket{0}+\beta\,\ket{1}\;, \nonumber\\
\ket{\Psi} &\stackrel{(1-x)/3}\longrightarrow & \sigma_1\sigma_3\ket{\Psi}=-
\alpha\,\ket{1}+\beta\,\ket{0} \;.
\end{eqnarray}
It is similar to the two-Pauli
channel considered previously\cite{T}, except the depolarizing   
channel flips the qubit with equal
probability for all three Pauli operators.

A general (input) state in the Bloch sphere representation
can be written as
\begin{equation}
\rho_x=\frac{1}{2}\Big(I + \vec{a}\cdot\vec{\sigma}\Big)\;.
\end{equation}
Here, $\vec{a}=(a_1,a_2,a_3)$ is the Bloch vector of length unity or less,
and $\vec{\sigma}$ is the vector of Pauli matrices.
The action of the channel on this density state is:
\begin{equation}
{\cal N}(\rho_x)=
\frac{1}{2}\Big(I + \vec{b}\cdot\vec{\sigma}\Big)\;,
\end{equation}
in which 
\begin{equation}
\vec{b}=( {\frac 4 3}x -{\frac 1 3})\Big(a_1,\,a_2,\,a_3\Big)\;.
\end{equation}
It is clear from this expression that the whole Bloch vector is shrunken
by a factor of $(4 x -1)/3$.

The logarithm of a density matrix is defined in its orthogonal basis, i.e.,
pure state decomposition\cite{AP}.
If $\rho_x$ can be written as 
\begin{equation}
\rho_x = \sum_i a_i \proj{i},
\end{equation} 
then the logarithm of $\rho_x$ is given by
\begin{equation}
\log \rho_x = \sum_i \log a_i \proj{i}.
\end{equation}
The von Neumann entropy can be shown to become
\begin{equation}
H (\rho_x) = - \sum_i a_i \log a_i,
\end{equation}
in which $a_i$s are the eigenvalues of the density matrix $\rho_x$.
An eligible density matrix never have negative or imaginary eigenvalue.

\begin{figure}
\epsfxsize=7.5cm\epsfbox{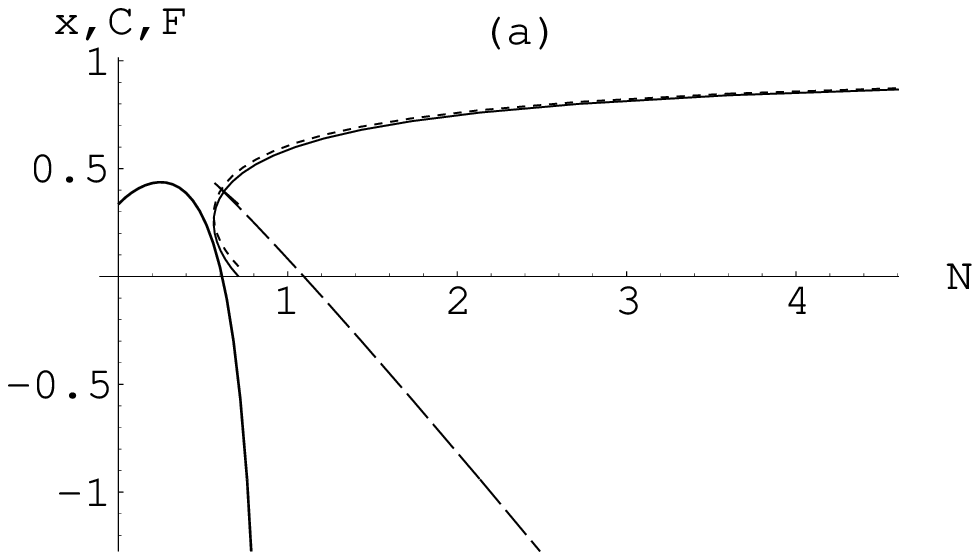}
\epsfxsize=7.5cm\epsfbox{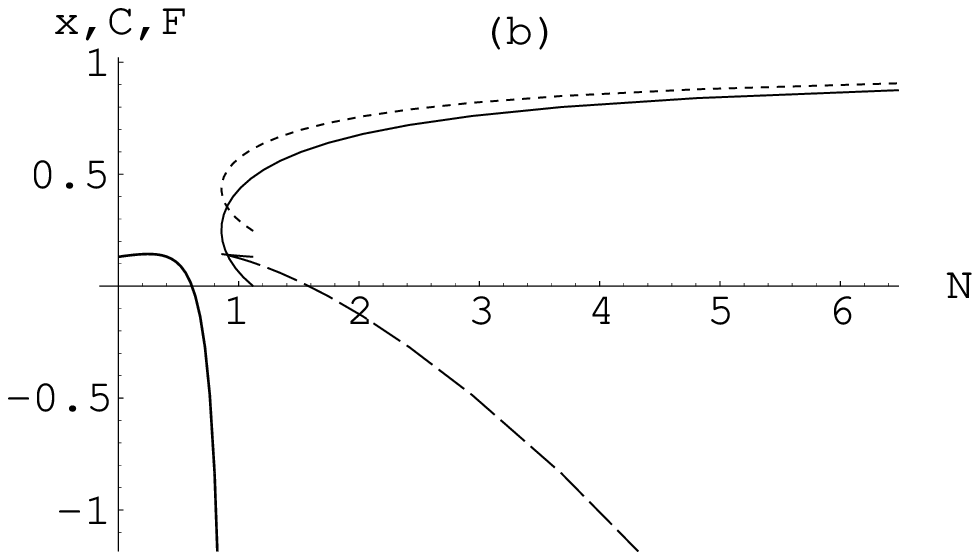}
\vskip 0.3cm
\caption{Parametric plots of the 
retention rate, $x$, versus noise, $N$ (solid lines);
coherence information, C, versus
noise, N, (long dashed lines);
fidelity, F, versus noise, N, (short dashed lines) and
coherent information, C, versus flipping rate, x, (thick solid line). for the parameter
$x$ from $0$ to $1$ and
initial states:
(a) $a_1 = 1/10, a_2 = 2/10, a_3 =3/10;$
(b) $a_1 = 8/10, a_2 = 3/10, a_3 =1/10$. 
}
\label{fig1}
\end{figure}

The matrix $W$ for the depolarizing channel read,
\begin{equation}
W=\left(\begin{array}{cccc}
x & a_1 \sqrt{\frac {x (1-x)} 3} & a_3  \sqrt{\frac {x (1-x)} 3}
& i a_2 \sqrt{\frac { x(1-x)} 3}\\
a_1 \sqrt{{\frac {x (1-x)} 3}} & {\frac {1-x} 3}&
{\frac {ia_2 (1-x)} 3} & {\frac {a_3 (1-x)} 3} \\
a_3 \sqrt{{\frac {x (1-x)} 3}} & 
{\frac {-ia_2 (1-x)} 3} & 
{\frac {1-x} 3}&
{\frac {a_1 (1-x)} 3} \\
- i a_2 \sqrt{{\frac { x(1-x)} 3}} & {\frac {a_3 (1-x)} 3}
& {\frac {a_1 (1-x)} 3} & {\frac {1-x} 3} \\
\end{array}\right).
\end{equation}
It has four eigenvalues, namely
$\lambda_{1,2} = (1-x)(1 \pm |a|)/3$ 
and 
$\lambda_{3,4}= \left[(1+2x) \pm \sqrt{12 x(x-1) (1-|a|^2)+(1+2x)^2} \right]/6$, in which
$|a|=\sqrt{a_1^2+a_2^2+a_3^2}$. Hence,
\begin{equation}
N  = - \sum_{i=1}^4 \lambda_i \log_2 \lambda_i,
\end{equation}
while
\begin{equation}
H (\rho_y) = - \sum_{i=1}^2 \theta_i \log_2 \theta_i,
\end{equation}
with $\theta_{1,2} = \left[ 3 \pm (1-4x) |a| \right]/6$.
This $x-N$, $C-N$ relationship is plotted in Fig.~\ref{fig1}.
Similar enhancements like the two-Pauli channel considered previously are
found\cite{T}. 
There are clear evidences of  noise enhancement for the fidelity.
In order to compare with previous authors 
the coherent information versus flipping rate they are also 
plotted in the figure.

Furthermore, for a communication channel
the (entangled) fidelity,
\begin{equation}
F = \sum_{\mu} (\tr \rho_x A_{\mu})(\tr \rho_x A_{\mu}^{\dagger}),
\end{equation}
is also of our concern, since it represent the quality of the signal 
transmitted. For the depolarizing channel
\begin{equation}
F = {\frac 1 3 }{ |a|^2  (1 - x)}  + x.
\end{equation}
We found not only for the capacity but also for for the noise and for 
the fidelity,
$a_1, a_2$ and $a_3$ are on the same footing.
The channel is symmetric for the exchange of $a_1$, $a_2$ and
$a_3$. Only the length of the vector $\vec a$ matters.
The relation between the
fidelity and the noise is  plotted with the coherent information
in Fig.~\ref{fig1}. 
Note that some people might think fidelity can be used as 
a measure of the noise strength. However, it is only an {\em indirect}
measure.
It measures the effect of the noise instead of the noise itself.

\section{conclusion}
In summary, a quantum stochastic resonance like scenario
in a quantum communication channel is considered. 
It is important to note here that which definition of a parameter
of the system is a suitable one is depending upon the problem one
would like to ask.
Within the context of stochastic resonance there is no right or wrong
of the parameters used.
In stochastic resonance we can look for any definition we like for
the resonance. In this paper we have demonstrated two.

Gisin and Wolf said quantum cryptography lies at the intersection
of two of the major sciences of the $20$th century: quantum mechanics and
information theory\cite{GW}.
The present work is at a three-junction of modern sciences, namely,
stochastic resonances, quantum mechanics, and information theories.

Other effects in quantum communication are under further consideration.

I thank 
Mrs. L. L. Harn, the librarian of NCHU physics department,
for providing me the computing facility and the library access.

\begin{references}
\bibitem{AS}{A.~Steane, {\em Rep. Prog. Phys.}, {\bf 61}, 117 (1998).}
\bibitem{S}{B.~ Schumacher, {\em Phys. Rev. A} {\bf 54}, 2614 (1996).}
\bibitem{DEJ}{D.~ Deutsch, A.~ Ekert, R.~ Jozsa, 
C.~ Macchiavello, S.~Popescu and
A.~Sanpera, {\em Phys. Rev. Lett.} {\bf 77}, 2818 (1996).}
\bibitem{BBP}{C.~H.~Bennett, G.~Brassard, S.~Popescu, B.~Schumacher,
J.~A.~Smolin and W.~K.~Wootters, {\em Phys. Rev. Lett.} {\bf 76}, 722 (1996).}
\bibitem{SH}{P.~W.~Shor,  in {\em Proceedings of the 35th Annual Symposium
on Foundations of Computer Science}, (IEEE Computer Society Press, New York, 
1994), p124.}
\bibitem{GC}{N.~A.~Gershenfeld and I.~L.~Chuang, {\em Science}, {\bf 275},
350 (1997).}
\bibitem{BG}{A.~Bulsara and L.~Gammaitoni, {\em Phys. Today}, {\bf  49}, 39
(1996).}
\bibitem{BZ}{A.~ R.~ Bulsara and A.~ Zador, {\em Phys. Rev. E}, {\bf 54}, 2185
(1996).}
\bibitem{FCB}{F.~ Chapeau-Blondeau, {\em Phys. Rev. E}, {\bf 55}, 2016 (1997).}
\bibitem{RABI}{J.~W.~C.~Robinson, D.~E.~Asraf,
A.~R.~Bulsara and M.~E. ~Inchiosa, {\em Phys. Rev. Lett.}, 
{\bf 81}, 2850 (1998).}
\bibitem{SS}{P.~W.~Shor and J.~A.~Smolin,
{\em quant-ph/9604006}.}
\bibitem{LL}{S.~Lloyd, {\em Phys. Rev. A}, {\bf 55}, R1613, 
quant-ph/9604015 (1997).}
\bibitem{T}{J.~J.-L.~Ting, {\em Phys. Rev. E}, {\bf 59} 2801 (1999).}
\bibitem{JP}{J.~Preskill, quant-ph/9904022.}
\bibitem{S96b}
{B.~Schumacher and M.~A. Nielsen, {\em Phys. Rev. A}, {\bf 54},  2629  (1996).}
\bibitem{BST}{H.~Barnum, J.~A.~Smolin  and B.~M.~Terhal,  
{\em Phys. Rev. A}, {\bf 58}, 3496 
(1998).}
\bibitem{BKN}{H.~Barnum, E.~Knill and M.~A.~Nielsen,      
{\em IEEE Trans. Info. Theory, quant-ph/9809010}, (1998).}
\bibitem{N}{J.~von Neumann, 
{\em Mathematical Foundations of Quantum Mechanics}, 
translated by E.~T.~Beyer (Princeton University Press, Princeton, 1955).}
\bibitem{AP}{A.~Peres, {\em Quantum Theory: Concepts and Methods},
(Kluwer Academic Publisher, 1993), page 68.}
\bibitem{GW}{N.~Gisin and S.~Wolf, {\em quant-ph/9902048}.}
\end{references}
\end{document}